\documentclass[12pt]{article}
\usepackage{graphicx}
\usepackage[bookmarksnumbered,colorlinks,plainpages]{hyperref}
\usepackage{amssymb,amsfonts,amsmath}
\usepackage[T2A]{fontenc}

\begin{document}

\title{The extrema of an action principle for dissipative mechanical systems}
\author{Tongling Lin and Qiuping A. Wang\thanks{Email: awang@ismans.fr} \\
{\small Laboratoire de Physique Statistique et Syst\`emes Complexes, ISMANS,} \\
{\small LUNAM Universit\'e, 44, Avenue, F.A. Bartholdi, 72000, Le Mans, France.} \\
{\small IMMM, UMR CNRS 6283, Universit\'e du Maine, 72085 Le Mans, France}}

\date{}

\maketitle

\begin{abstract}
A least action principle for damping motion has been previously proposed with a Hamiltonian and a Lagrangian containing the energy dissipated by friction. Due to the space-time nonlocality of the Lagrangian, mathematical uncertainties persist about the appropriate variational calculus and the nature (maxima, minima and inflection) of the stationary action. The aim of this work is to make numerical simulation of damped motion and to compare the actions of different paths in order to get evidence of the existence and the nature of stationary action. The model is a small particle subject to conservative and friction forces. Two conservative forces and three friction forces are considered. The comparison of the actions of the perturbed paths with that of the Newtonian path reveals the existence of extrema of action which are minima for zero or very weak friction and shift to maxima when the motion is overdamped. In the intermediate case, the action of the Newtonian path is neither least nor most, meaning that the extreme feature of the Newtonian path is lost. In this situation, however, no reliable evidence of stationary action can be found from the simulation result. 
\end{abstract}

in classical mechanics) 

\vspace{2 cm}

\section{Introduction}

The Least Action Principle (LAP)\cite{Maupertuis}-\cite{Goldstein}, or the variational principle in general, for damped motion of dissipative mechanical systems is a longstanding question\cite{Sieniutycz,Vujanovic}. To our knowledge, the first application of variational calculus to damped motion dates back to Euler's calculus in 1744 for the brachistochrone (shortest time) problem with friction\cite{Goldstine}. More recently, much effort has been focused on finding a Lagrangian function, or a LAP with vanishing first variation of the action to derive the equation of motion for dissipative systems\cite{Sieniutycz}-\cite{Schuch}, including the proposition by Rayleigh of a `dissipative function' $D=\frac{1}{2}\zeta \dot{x}^2$ to write $\frac{d}{dt}\left(\frac{\partial L}{\partial \dot{x}}\right)+\frac{\partial D}{\partial \dot{x}}-\frac{\partial L}{\partial x}=0$, where $\zeta$ is the drag constant in the Stokes law  $\vec{f}_d=-m\zeta \vec{\dot{x}}$ and $m$ the mass of the damped body. Although the equation of motion is kept in a similar form as Lagrangian equation, LAP is not recovered since there is no a single Lagrangian for defining an action with vanishing first variation. Other major propositions include the Bateman approach\cite{Bateman} to introduce complementary variables and equations, the definition of dissipative Lagrangian by multiplying the non dissipative one with an exponential factor $exp(\zeta t)$\cite{Sanjuan} where $t$ is the time, the fractional derivative formulation\cite{Riewe}, and the pseudo-Hamiltonian mechanics\cite{Duffin} where a parameter was introduced to characterize the degree of dissipation. The reader is referred to the reviews in \cite{Sieniutycz,Vujanovic,Gray,Riewe,Duffin} about the details of these propositions.

These solutions have considerably contributed to the development of variational calculus for dissipative mechanical systems. Nevertheless, some drawbacks persist. We can cite for instance the limited application to only some special systems and frictions, the non-uniqueness of the Lagrangian, the lack of energy connection of the Lagrangian and Hamiltonian, and the lose of the optimal character of action in general\cite{Sieniutycz,Vujanovic,Gray}. In order to avoid these inconveniences and to recover the original elegance of the LAP, it is necessary to find a unique and universal Lagrangian function with close energy connection and without restriction to specific frictions for defining an generic action by time integral of that Lagrangian. The solution we have proposed\cite{Wang} is a Lagrangian constructed for an isolated total system which contains a moving body and its environment which are coupled to each other by friction. Although the damped body is a nonconservative system, the energy of the total system is conserved since energy is only transfered from the body to the environment. It can be expected that the macroscopic smooth motion (no thermal fluctuation considered) of this total Hamiltonian system may be subject to a formalism of the LAP. The key step of the formulation was, for defining the action $A=\int Ldt$, to find a single and unique Lagrangian function $L$ with close energy connection as the usual one (i.e something like $L=K-V$ with kinetic energy $K$ and potential energy $V$). This is not that easy because a part of $K$ and $V$ of the moving body is transformed by the friction into other energy forms such as heat, acoustic or electromagnetic waves. It is still unknown how to include these forms of energy in the Lagrangian which, without considering this part of energy, may lose its kinship with the Hamiltonian $H$, e.g., the Legendre transformation (see below) which plays an important role in the usual LAP formulation of Hamiltonian/Lagrangian mechanics. 

In our proposition\cite{Wang}, this is solved thanks to a potential like expression of the relation between the friction force and the dissipated energy : the derivative of the latter with respect to the instantaneous position of the damped body yields the friction force. This Lagrangian could also be derived from the virtual work principle\cite{Wang}. In this formulation of LAP for dissipative systems, the three major conventional formulations of Hamiltonian/Lagrangian mechanics, i.e., the Lagrangian, the Hamiltonian and the Hamilton-Jacobi equations, together with the Legendre transformation, are all preserved. 

However, as will be shown below, the nonlocal expression of the dissipative energy makes the Hamiltonian and Lagrangian non local in space and time, which yields an action functional with double time integral. This leaves an uncertainty about how to use the usual variational calculus which has always been applied to actions with local Lagrangian. Another matter of investigation is about the nature (maximum, minimum or inflection) of the possible stationarity of action. If the action of the optimal path is a minimum when there is no friction, does this minimum survive energy dissipation? If not, what will be the nature of the action stationarity? 

In what follows, we will discuss in detail the problem coming from this non locality and describe the numerical simulation of damped motion as well as the results. The purpose of this simulation is to calculate the actions along the optimal path and many other variational paths created with tiny perturbations of the optimal one, and to see, by comparing the values of these actions, whether there are traces of stationary action and of the nature of the stationarity.

\section{LAP for damped motion}

LAP was originally formulated for Hamiltonian system, i.e., the Hamiltonian function $H=K+V$ satisfies the Hamiltonian equations\cite{Arnold}. The real trajectories between two given configuration points are prescribed by the LAP, a vanishing first variation $\delta A$ created by tiny perturbation of the trajectories\cite{Lanczos}:

\begin{equation}    \label{e1}
\delta A=0
\end{equation}
where the action $A=\int_0^{T} Ldt$ is a time integral of the Lagrangian $L=K-V$ on the trajectory from a point $a$ to a point $b$ over a fixed time period $T$ (suppose $t_a=0$ and $t_b=T$). One of the important results of this variational calculus is the Euler-Lagrange equation given by\cite{Lanczos} (for one freedom $x$)

\begin{equation}    \label{e2}
\frac{d}{dt}\left(\frac{\partial L}{\partial \dot{x}}\right)-\frac{\partial L}{\partial x}=0
\end{equation}
where $\dot{x}$ is the velocity. In many cases when $H$ and $L$ do not depend on time explicitly, a Hamiltonian system is energy conservative. However, for damped motion with friction force $f_d$, the above equation becomes $\frac{d}{dt}\left(\frac{\partial L}{\partial \dot{x}}\right)-\frac{\partial L}{\partial x}=f_d$ which is equivalent to write $\int_0^{T} (\delta L +f_d\delta x)dt=0$, meaning that Eq.(\ref{e2}), with a unique single Lagrangian function defining $A$, is lost.

The solution\cite{Wang} we proposed is to consider a whole system including the damped body and its environment coupled to each other by a dissipative force. This whole system has a total Hamiltonian $H=K+V+H_i+H_e$ at time $t$ where $H_i$ is the interaction energy between the moving body and its environment, $H_e=H_0+E_d$ is the total energy of the environment, $H_0$ is its energy at the initial moment of the motion and $E_d$ the energy dissipated by the friction force from the body to the environment from the initial moment ($t=0$) to the present moment $t$. Since the variation only concerns the period ($0\leq t \leq T$), $H_0$ is a constant and can be dropped from the variational calculus. On the other hand, the energy $H_i$, responsible for the friction law, is determined by the coupling mechanism between the moving body and the parts of the environments that are the closest to the body-environment interface. If this interface (body's shape and size, body-environment distance, nature of the closest parts of the environment etc) and the friction law do not change significantly during the considered period, the interaction mechanism should not change with the virtual variation of paths. So $H_i$ can also be neglected in the calculus. Consequently, for the variational calculus, we only consider the effective Hamiltonian $H=K+V+E_d$ and Lagrangian $L=K-V-E_d$\cite{Wang}. $E_d$ is given by the negative work of the friction force $\vec{f}_d=-f_d\vec{k}$ from $a$ to a position $x(t)$ along the path $s=s(t)$  ($0\leq t \leq T$) where $\vec{k}$ is a unitary vector indicating the direction of the motion at a point $s(t)$ and the magnitude of the friction force $f_d$ can be any function of time, position and velocity. It reads

\begin{eqnarray}                                            \label{e3}
E_d &=&-\int_{x_a}^{x(t)}\vec{f}_d\cdot d\vec{s}(\tau)=\int_0^{x(t)}f_d\vec{k}\cdot ds(\tau)\vec{k}\\ \nonumber &=&\int_0^{x(t)}f_d\cdot ds(\tau)=\int_0^tf_d(\tau)\dot{s}(\tau)d\tau
\end{eqnarray}
where $\tau$ is any time moment between $t_a=0$ and $t$, $d\vec{s}(\tau)=\vec{\dot{s}}(\tau)d\tau$ a small displacement along $s$ at time $\tau$ with $\dot{s}(\tau)=ds(\tau)/d\tau$. $E_d$ depends on both the past trajectory and the present instantaneous position $x(t)$. Hence Eq.(\ref{e3}) is a space-time nonlocal expression of $E_d$. The action has been given by\cite{Wang}

\begin{equation}                                            \label{e4}
A=\int_0^{T}(K-V-E_d)dt.
\end{equation}
Due to the non locality of $E_d$, both $H$ and $L$ defined above are non local. The variational calculus will in this case need more attention than with the usual local Lagrangian. 

\begin{figure}[ht]
\centering
  \includegraphics[width=0.7\linewidth]{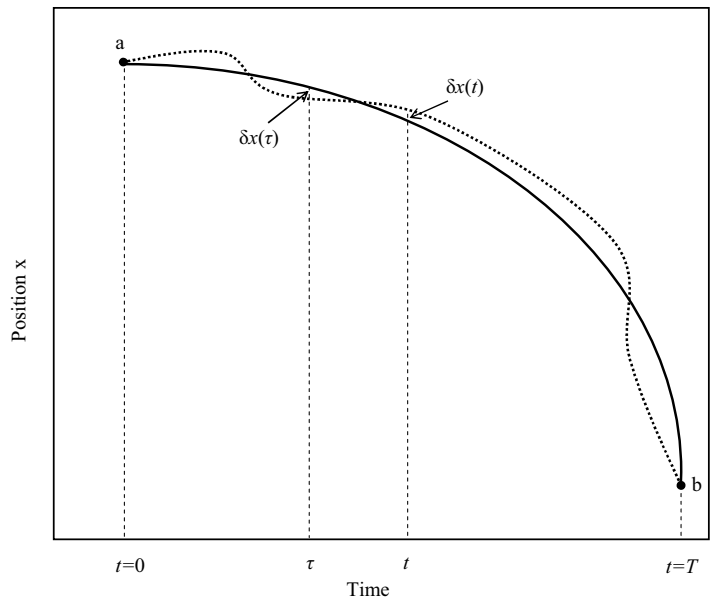}
\caption{Illustration of an exaggerated variation operation over the entire optimal path (thick line) from the point $a$ to the end point $b$.}
\label{Fig1}
\end{figure}

Fig.\ref{Fig1} illustrates a variation operation over the entire optimal path (thick line) from the point $a$ to the end point $b$. Let $\delta(t)$ be the variation on the position at time $t$, with $\delta(a)=\delta(b)=0$. In the conventional calculus, the local Lagrangian undergoes only a variation at $t$ produced by the position variation $\delta x(t)$ in the Lagrangian, i.e., $\delta A=\int_0^{T} [L(x+\delta x(t),\dot{x}+\delta\dot{x}(t),t)-L(x,\dot{x},t)]dt$. The whole variation over the entire path is then taken into account through the time integral from $a$ to $b$. Now with the non local action of Eq.(\ref{e4}), it seems logical to consider the variation $\delta x(\tau)$ at the moment $\tau$ before $t$ since it may change $E_d$.

This calculus, using both $\delta x(t)$ and $\delta x(\tau)$, is shown in the Appendix. It results in a wrong equation of motion. Two reasons are possible for this failure: either $\delta A =0$ is not true, or the calculus is not appropriate. One may prefer the calculus in view of its mathematical rigor. But a doubt about its physical validity can arise. Indeed, considering $\delta x(\tau)$ means considering the influence of a prior event at time $\tau$ on a later motion at time $t$. Naturally, if the path at $\tau$ is changed a little bit, the work of the friction force will change. This variation of dissipative energy at a previous moment may produce variation of the energy at a later moment if the conservation of energy of the total system is taken into account as a constraint of the variational calculus. However, energy conservation is not a constraint in this version of LAP using the action defined with Lagrangian\cite{Arnold,Gray2}, meaning that the consideration of the variation at time $\tau$ to derive the equation of motion of a later moment $t$ is questionable. 

In reality, the energy of the total system is conserved, hence $H=K+V+E_d$ is a constant. This fact can indeed be used as a constraint of variational calculus, as has been done in the usual calculus using Maupertuis action instead of Lagrangian action\cite{Arnold,Gray2}. We have followed this line in \cite{Wang} and shown that the minimization of the Maupertuis action $A_M=\int_a^bpdx=\int_a^b2Kdt$ ($p$ is momentum $p=m\dot{x}$) gave rise to the correct equation of damped motion \cite{Maupertuiscalculus}
\begin{equation}                                            \label{e5}
m\ddot{x}=-\frac{\partial (V+E_d)}{\partial x}=-\frac{\partial V}{\partial x}-m\zeta\dot{x}.
\end{equation}
There is no problem of non locality because $p$ and $K$ are local and the variation $\delta K=-\delta (V+E_d)$ only takes place at $t$. 

It is well known that the least Maupertuis action $\delta A_M=0$ with the constraint of energy conservation is equivalent to the least Lagrange action $\delta A=0$ with the constraint of constant duration of motion, as long as the Legendre transformation $L=p\dot{x}-H$ is valid\cite{Gray2}. The reasoning is straightforward. From the definition of $A$ and $A_M$ and the Legendre transformation, we have $A=A_M-\bar{H}T$ and the following variational calculus $\delta A+\bar{H}\delta T=\delta A_M- T\delta\bar{H}$ where $\bar{H}=\frac{1}{T}\int_0^THdt$ is the time average of the Hamiltonian along the path. This variational relation implies that $\delta A_M=0$ with energy conservation along the path $\delta\bar{H}=0$ is equivalent to the LAP $\delta A=0$ with fixed duration of motion $\delta T=0$. This equivalence means that $\delta A_M=0$ and $\delta A=0$ should correspond, under different constraints, to the same equation of motion. This is one of the elements that advocate in favor of the LAP for dissipative systems with $A=\int_0^TLdt$ where $L=K-V-E_d$ can be derived from the Lengendre transformation $L=p\dot{x}-H$. 

Coming back to the calculus with non local Lagrangian, without energy conservation as constraint, a prior variation of path and the concomitant energy change cannot influence later variation of energy and action. From physical point of view, the energy $E_d$ has already been transformed into other forms of energy in the surroundings. Its variation prior to $t$ can modify the surrounding's motion but not the instantaneous motion of the body at $t$ if the feedback from the surrounding onto the body (heating, acoustic or electromagnetic chocks) is neglected as we have assumed\cite{Wang}.

For the above reasons, we have considered only the variation at $t$ and derived Eq.(\ref{e5}) from $\delta A=0$. It was also shown that this approach was supported by the virtual work principle and by a differential version of LAP\cite{Wang}. The argument of the differential LAP is the following. If $A$ is a minimum over the entire optimal path between, the same must be true over any infinitesimal segment of the trajectory, i.e., the time integral of $L$ over a tiny segment must be a minimum whatever its length is. If not, we can always play with this segment to make $A$ smaller than its minimal value along the optimal path. One can therefore choose any segment on the optimal path and make a small variation of it $\delta x(t)$, without making variation elsewhere. This operation avoids the problem of a prior variation before or after $t$, and obviously leads to Eq.(\ref{e5}).

Summarizing the above discussion, it seems physically reasonable to remove the variation of the action $A$ due to $\delta x(\tau)$ in spite of the seemingly rigorous calculus considering $\delta x(\tau)$ presented in the Appendix. A possible technique of verification of this choice is to make numerical simulation of damped motion and to calculate the action $A$ along the optimal path given by Newtonian equation and many other perturbed paths around the optimal one. Since the perturbations are of arbitrary magnitude, the perturbed paths can be considered as paths undergoing variations. The comparison of the values of the action $A$ should reveal whether or not it is likely for the optimal path to have action extrema, what is their nature (maximum, minimum or inflection) and how they evolve with dissipation. A priori, we can do this for both $A$ and $A_M$. However, the variation of $A_M$, as mentioned above, needs the constraint of fixed Hamiltonian and the arbitrary duration of motion $T$ between two given points $a$ and $b$. This constraint is more difficult to produce in the numerical simulation than with fixed $T$ and arbitrary Hamiltonian for the variation with $A$. In this latter case, the perturbation is just created randomly at each step of the motion. The number of steps of the discrete motion is the same for all the paths, which is easy to produce with iterative computation. Hence the action $A$ defined with the Lagrangian $L=K-V-E_d$ is used throughout this work.

\section{Optimal path and action with constant force and Stokes' drag}
\begin{figure}[ht]
\centering
  \includegraphics[width=0.7\linewidth]{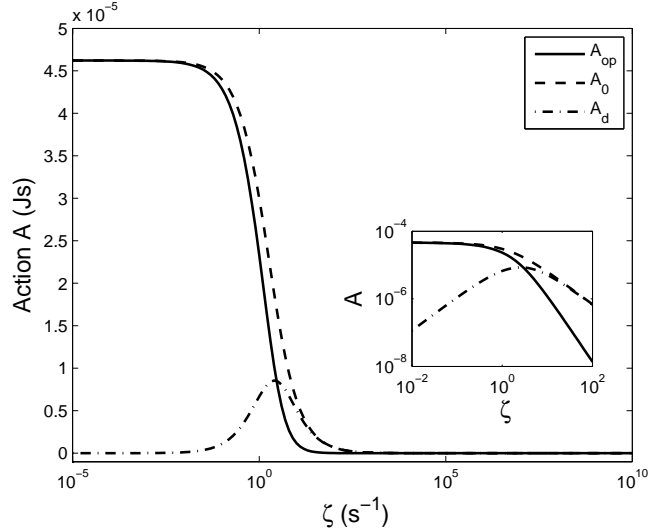}
\caption{$\zeta$ dependence of the actions for the optimal path with $T=1$ $s$ (for $n_s=1000$ steps with $10^{-3}$ $s$ each step). $A_{op}=A_0-A_d$ is the optimal action (solid line), $A_{0}$ is the usual action (dashed line), $A_{d}$ is the dissipative part of the action (dot dashed line). The drop point $\zeta_c$ can be estimated by $\zeta_c T=1$. The inset is a zoom of the zone around $\zeta_c$ in double logarithm plot.}
\label{Fig2}
\end{figure}

\begin{figure}[ht]
\centering
  \includegraphics[width=0.7\linewidth]{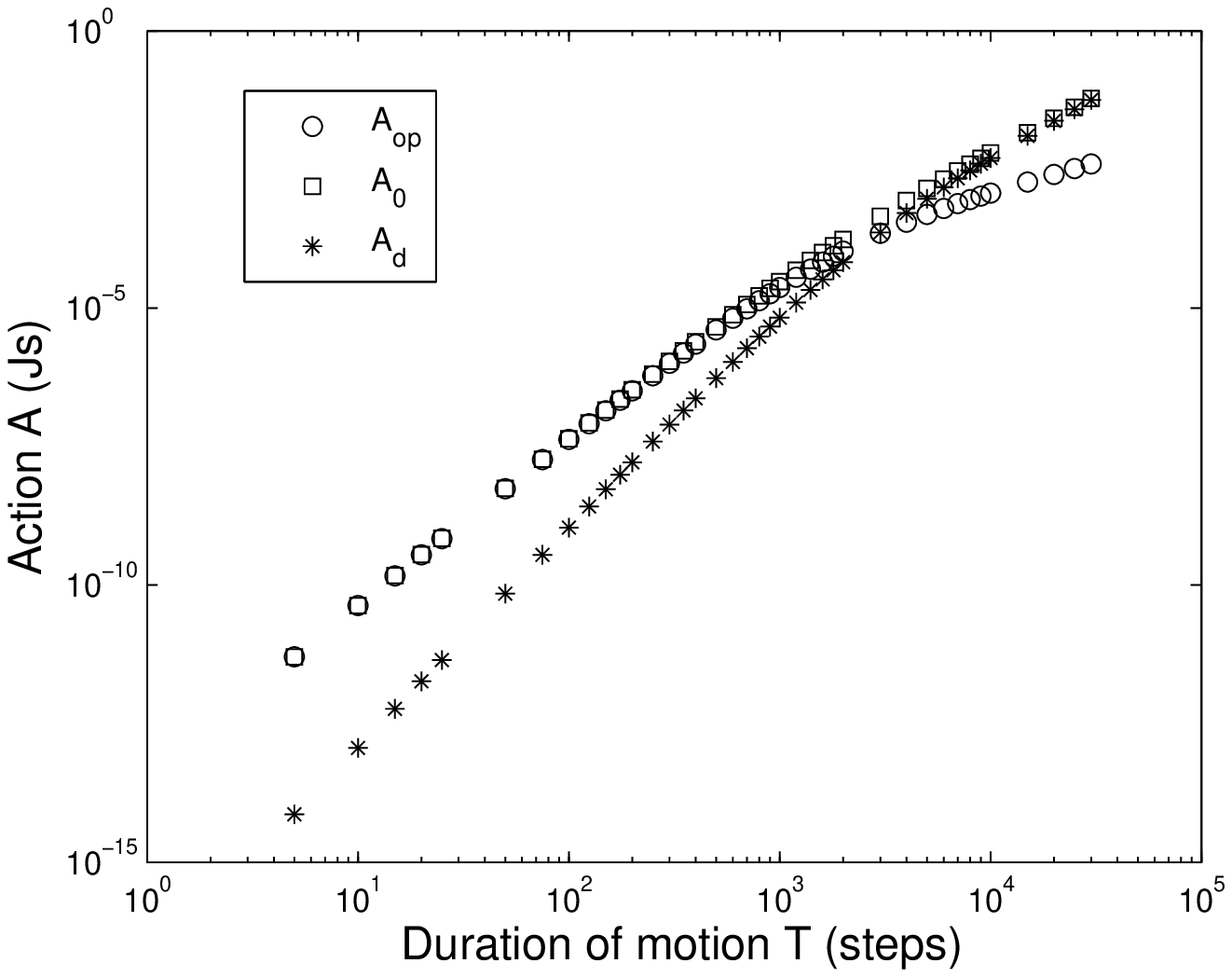}
\caption{$T$ dependence of the actions for the optimal path with $\zeta=1$ $s^{-1}$, where $A_{op}=A_0-A_d$ is the optimal action (circles), $A_{0}$ is the usual action (squares), $A_{d}$ is the dissipative part of the action (stars).}
\label{Fig3}
\end{figure}

The first case we consider is a small particle of mass $m=1.39\times10^{-6}$ kg subject to a constant force $f=mg$ where $g=10$ $ms^{-2}$. The friction is given by the Stokes' drag, i.e., $f_d=m\zeta \dot{x}$, where $\zeta$ is the drag constant. The optimal path corresponding to $\delta A=0$ or given by Eq.(\ref{e5}) is $x(t)=\frac{g}{\zeta^2}(1-e^{-\zeta t})-\frac{g}{\zeta}t$ for $x(0)=0$ and $\dot{x}(0)=0$. The optimal action $A_{op}=\int_0^{T}(\frac{m}{2}\dot{x}^2-mgx-m\zeta\int_0^t\dot{x}^2d\tau)dt$ can be calculated analytically :

\begin{equation}                                            \label{e6}
A_{op}=\frac{mg^2}{\zeta^2}(-\frac{1}{2\zeta} e^{-2\zeta T}+\frac{2}{\zeta} e^{-\zeta T}-\frac{3}{2\zeta}+T)
\end{equation}
whose $\zeta$ and $T$ dependence are shown in Fig.\ref{Fig2} and Fig.\ref{Fig3}, respectively. The usual action $A_0=\int_0^{T}(\frac{m}{2}\dot{x}^2-mgx)dt$ is given by
\begin{equation}   																					\label{e7}
A_0=\frac{mg^2}{\zeta^2}(-\frac{1}{4\zeta} e^{-2\zeta T}+\frac{1}{4\zeta}-\frac{1}{2}T+\frac{1}{2}\zeta T^{2}).
\end{equation}
For $\zeta=0$, $A_0=\frac{1}{3}mg^2T^3$ as expected. If we define $A_d=\int_0^{T}E_d dt=m\zeta\int_0^{T}\int_0^t\dot{x}^2d\tau dt$ as the dissipative action, it is given by
\begin{equation}																							\label{e8}
A_d=\frac{mg^2}{\zeta^2}(\frac{1}{4\zeta} e^{-2\zeta T}-\frac{2}{\zeta} e^{-\zeta T}+\frac{7}{4\zeta}-\frac{3}{2}T+\frac{1}{2}\zeta T^{2})
\end{equation}
which becomes $A_d\approx\frac{1}{12}mg^2\zeta T^4$ for small $\zeta$ and tends to zero for $\zeta\rightarrow 0$ when $T$ is finite and fixed. Notice that $A_{op}=A_0-A_d$. When $\zeta$ and $T$ are sufficiently small, so that $A_d$ can be negligible, $A_{op}=\frac{1}{3}mg^2T^3=A_0$. 

For large $\zeta$ ($10^{4}$ $s^{-1}$ for example as the particle is in glycerin at ambient conditions) and moderate $T$ (larger than, say, 1 $s$), the actions become $A_0\approx\frac{mg^2}{\zeta^2}(-\frac{1}{2}T+\frac{1}{2}\zeta T^{2})$, $A_d\approx\frac{mg^2}{\zeta^2}(-\frac{3}{2}T+\frac{1}{2}\zeta T^{2})$ and $A_{op}\approx\frac{mg^2}{\zeta^2}T$, which all decrease with increasing $\zeta$ and increase with increasing duration of motion $T$. $A_{op}$, $A_0$ and $A_d$ are numerically calculated along the optimal path. The particle moves from the initial point to the final point during the time interval $T=n_s\delta t=1$ $s$ where $n_s=1000$ is the number of steps and $\delta t=10^{-3}$ $s$ is the time increment of each step. The results are shown in Fig.\ref{Fig2} and Fig.\ref{Fig3}. The sharp drop in $A_{op}$ and $A_0$ is first due to the increase of $A_d$ (before its maximum) around $\zeta=1$ $s^{-1}$ and then to the decrease of the velocity $\dot{x}(t)=\frac{g}{\zeta}(e^{-\zeta t}-1)$ with increasing $\zeta$ for given $t$. The drop point $\zeta_c$ can be estimated by $\zeta_c T=1$, as expected from the exponential factors in Eqs.(\ref{e6}) to (\ref{e8}). The reader will find later that this point can be seen as a characteristic point in the change of nature of the extrema of action.

\section{Transition of extrema of action}

\begin{figure}[ht]
\centering
  \includegraphics[width=0.7\linewidth]{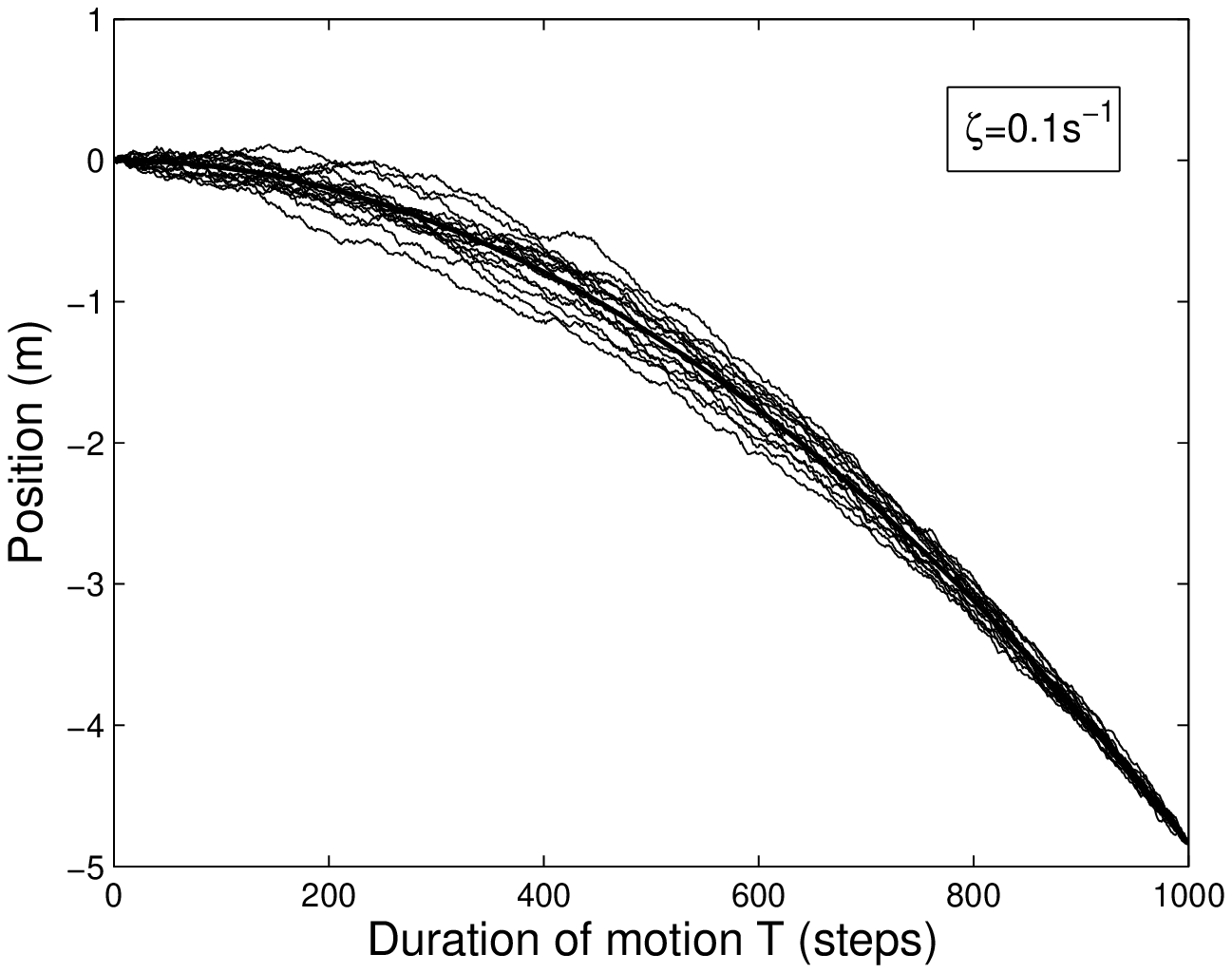}\\
  \caption{Samples of the different paths created randomly around the optimal path (thick line) given by the solution of Eq.(\ref{e5}) for a small particle moving between two fixed points in linear potential (constant force) and a viscous medium with Stokes'  drag constant $\zeta=0.1$ $s^{-1}$. The motion lasts $T=1$ $s$ with $n_s=1000$ steps and $\delta t=10^{-3}$ $s$ each step.}
  \label{Fig4}
\end{figure}

At this stage, it is not yet clear whether the vanishing first variation $\delta A=0$ yields a minimum, maximum or an saddle point action $A_{op}$. We know that when $A_{op}\approx A_0$ or $A_d\rightarrow 0$, the optimal action $A_{op}$ is a least one in this case of linear potential. The question is whether this minimum holds for any $\zeta$ and $T$ and how eventually it changes with these parameters. We propose in this work to investigate this matter by comparing the actions calculated along a large number of paths created by arbitrary variation from the optimal one. In the simulation algorithm, the arbitrary variation of position is made at each step of the motion by using a Gaussian distributed random displacements superposed on the optimal path $x(t)$ according to $x'_{i}=x'_{i-1}+\chi_i+x(t_i)-x(t_{i-1})$ where $\chi_i$ is the Gaussian random displacement at the step $i$ and $i=1,2 ...n_s$. A perturbed path is then a sequence of variations of position $\{x'_{0},x'_{1},x'_{2}\cdots ,x'_{n_s}\}$. The magnitude of the perturbation of position at each step can be controlled by the standard deviation $\sigma$ of the Gaussian distribution. Vanishing perturbation of the optimal path can be obtained with vanishing $\sigma$. Examples of these perturbed paths can be seen in Fig.\ref{Fig4}). These paths are sufficiently smooth and their number of steps $n_s$ is sufficiently large in order to calculate reliable velocity, energy, action and dissipative energy etc. The actions are calculated with different damping coefficient, duration of motion and $\sigma$ to see how the stationarity of action evolves with these parameters. 

\begin{figure}[ht]
\centering
\begin{minipage}[t]{0.49\linewidth}
  \includegraphics[width=0.9\linewidth]{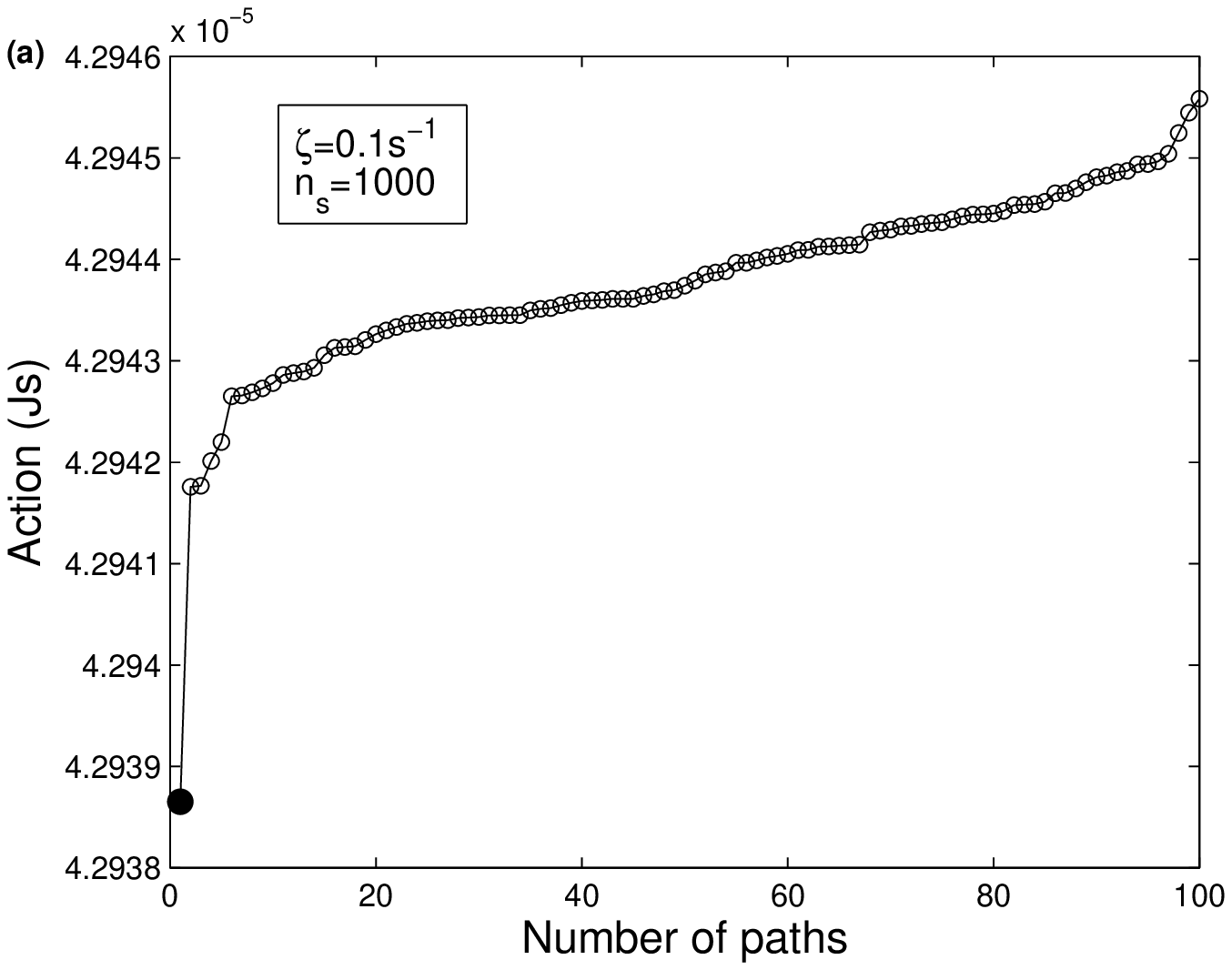}
\end{minipage}
\begin{minipage}[t]{0.49\linewidth}
  \includegraphics[width=0.9\linewidth]{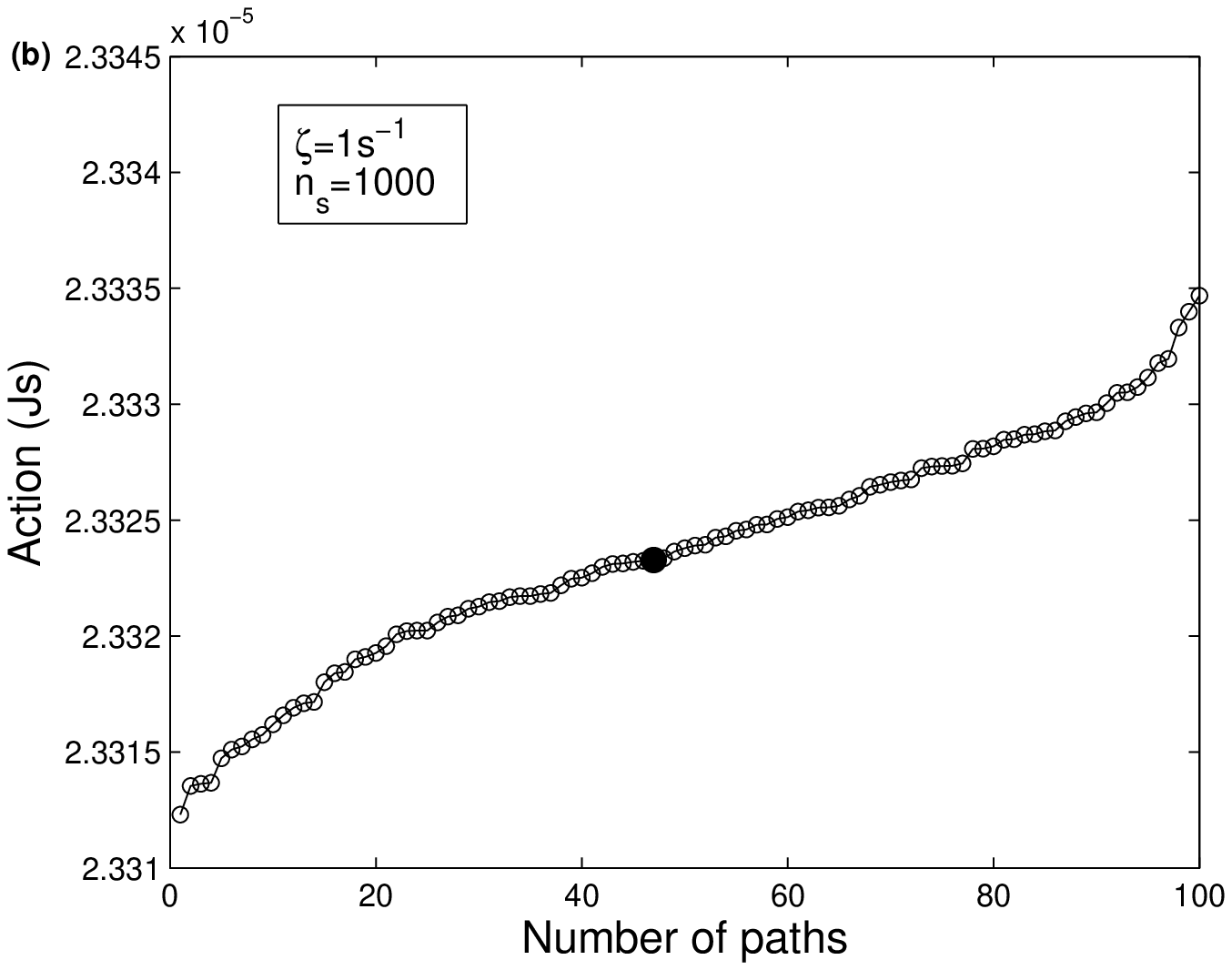}
\end{minipage}
\begin{minipage}[t]{0.49\linewidth}
  \includegraphics[width=0.9\linewidth]{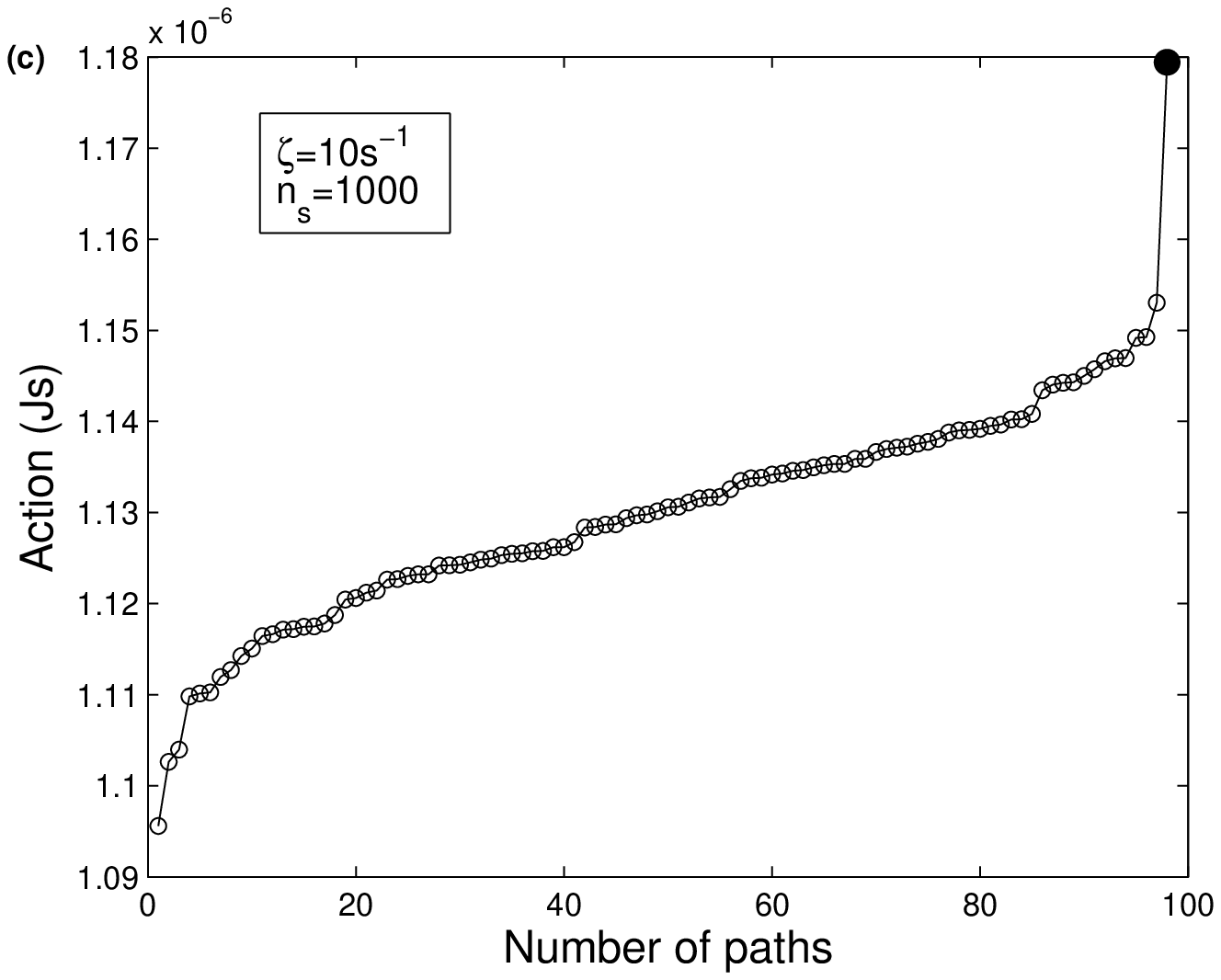}
\end{minipage}
\begin{minipage}[t]{0.49\linewidth}
  \includegraphics[width=0.9\linewidth]{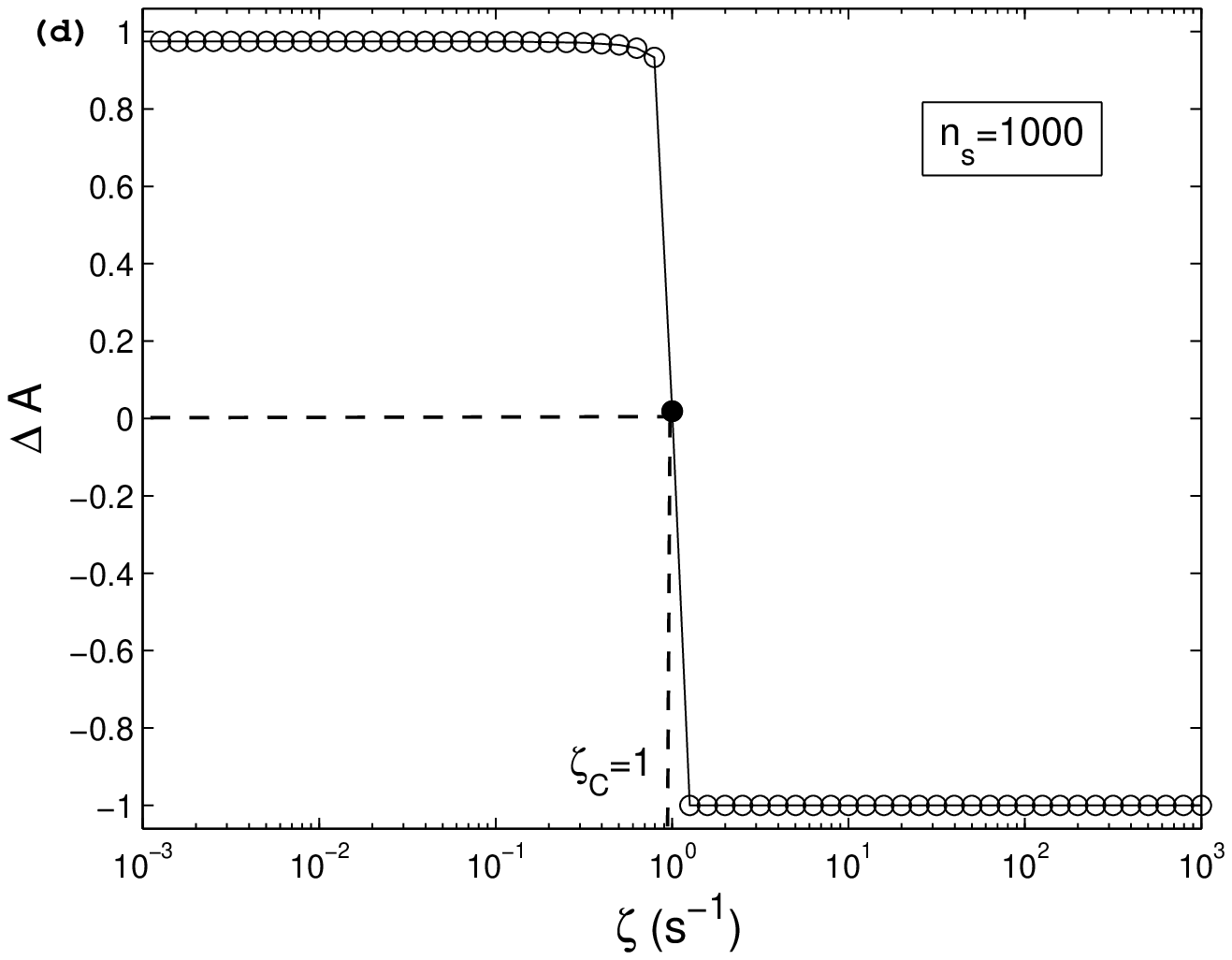}
\end{minipage}
  \caption{Illustration of the transition of extrema by comparison of the action of the optimal path (dots) with the actions of other paths (circles) created by random perturbation of optimal one. The number of steps is $n_s=1000$ with $\delta t=10^{-3}$ $s$ each step ($T=1$ $s$). (a) For $\zeta=0.1$ $s^{-1}$, (b) for $\zeta=1$ $s^{-1}$ and (c) for $\zeta=10$ $s^{-1}$. All calculations were made with an amplitude of variation $\sigma=0.1$ $mm$ for a total displacement of about 5 $m$ during $T$. (d) $\zeta$ dependence of the quantity $\Delta A=\frac{\bar{A}-A_{op}}{\left|\bar{A}\right|+\left|A_{op}\right|}$ where $\bar{A}$ is the average action over all the paths. $\Delta A$ can be used to characterize the evolution of extrema of $A$ in three regimes: the minimum regime ($\Delta A>0$), the maximum regime ($\Delta A<0$) and the saddle point regime around $\Delta A=0$ for $\zeta_c T\approx 1$.}
  \label{Fig5}
\end{figure}

In each simulation with given $\zeta$, $T$ and $\sigma$, we create about 100 paths and calculate their actions. An example of the comparison of the actions is shown in Fig.\ref{Fig5} (a), (b) and (c) for $\zeta=0.1$, $\zeta=1$ and $\zeta=10$, respectively. The duration of motion is $n_s=1000$ steps with $\delta t =10^{-3}$ $s$ each step ($T=1$ $s$). In (a), the optimal path (dot) has the smallest action $A_{op}$ among all the created paths (circles). In (b) $A_{op}$ is in the middle action range. In (c) $A_{op}$ becomes the largest action. So as $\zeta$ increases, there is an obvious transition of the stationary $\delta A=0$ from minimum regime (a) to maximum regime (c) in passing by a saddle point regime (b). In order to be sure of these results, we repeated the simulation with different values of $\zeta$ (from 0 to $10^{10}$ $s^{-1}$), $T$ (from $10^{-4}$ to 100 $s$) and $\sigma$ (from $10^{-10}$ to $10^{-3}$ m). The results are similar to those of Fig.\ref{Fig5}.

\begin{figure}
\centering
 \includegraphics[width=0.7\linewidth]{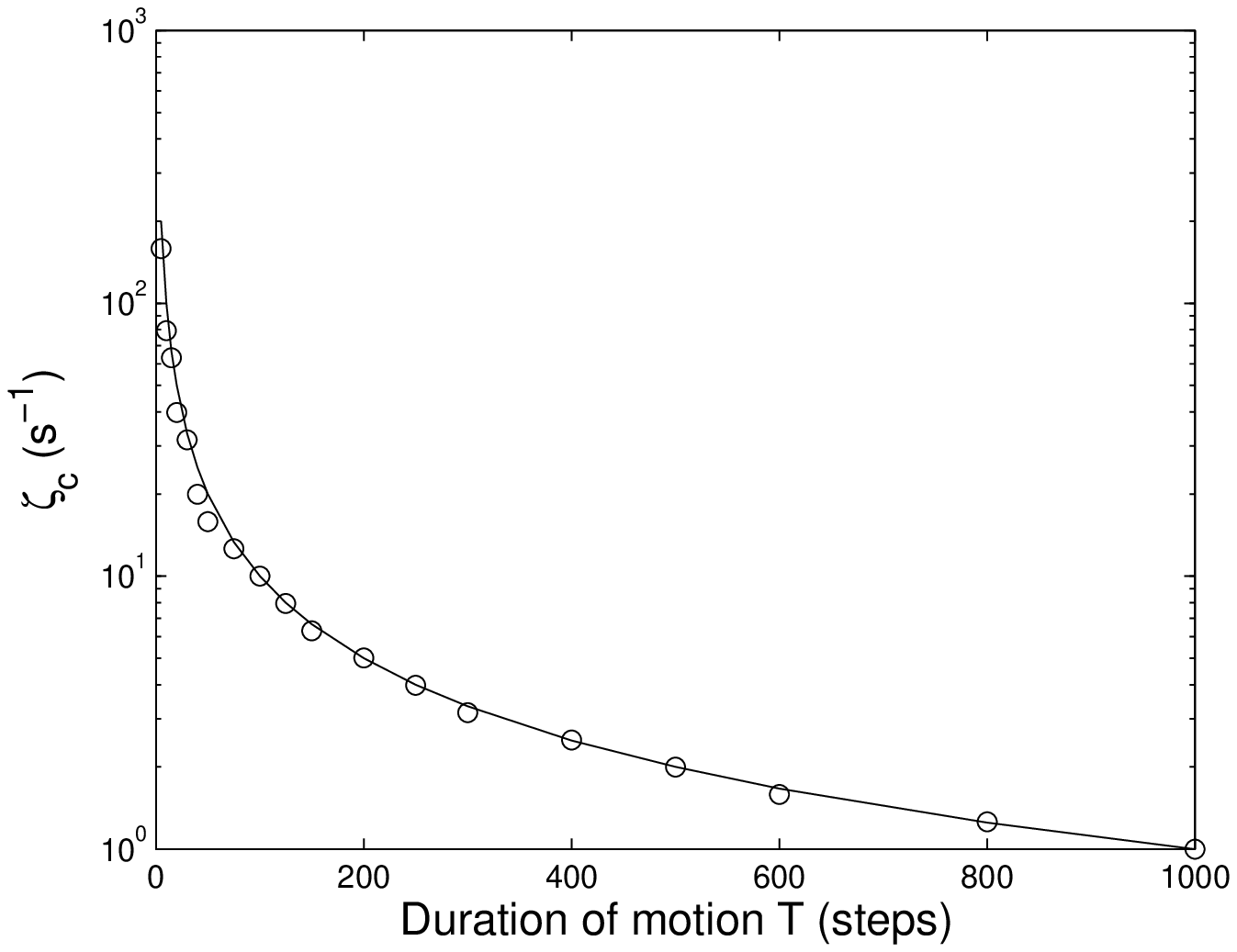}
  \caption{$T$ dependence of the characteristic value $\zeta_c$ which decreases with increasing $T$. It can be approximated by $\zeta_c T=1$.}
  \label{Fig6}
\end{figure}

To characterize this evolution, a quantity $\Delta A=\frac{\bar{A}-A_{op}}{\left|\bar{A}\right|+\left|A_{op}\right|}$ is defined where $\bar{A}$ is the average action over all the paths. This quantity is positive when $A_{op}$ is smaller than $\bar{A}$, negative when $A_{op}$ is larger than $\bar{A}$, and zero when $A_{op}$ is equal to $\bar{A}$. Fig.\ref{Fig5}(d) shows the $\zeta$ dependence of $\Delta A$ which can be characterized by a  point $\zeta_c$ which is determined by $\bar{A}=A_{op}$. The $T$-dependence of $\zeta_c$ is depicted in Fig.\ref{Fig6}. It can be approximated by $\zeta_c T=1$. Hence $A_{op}$ is in the minimum (maximum) regime for $\zeta$ much smaller (larger) than $\zeta_c$, and in the saddle point regime for $\zeta\approx\zeta_c$.

\begin{figure}
\centering
 \includegraphics[width=0.7\linewidth]{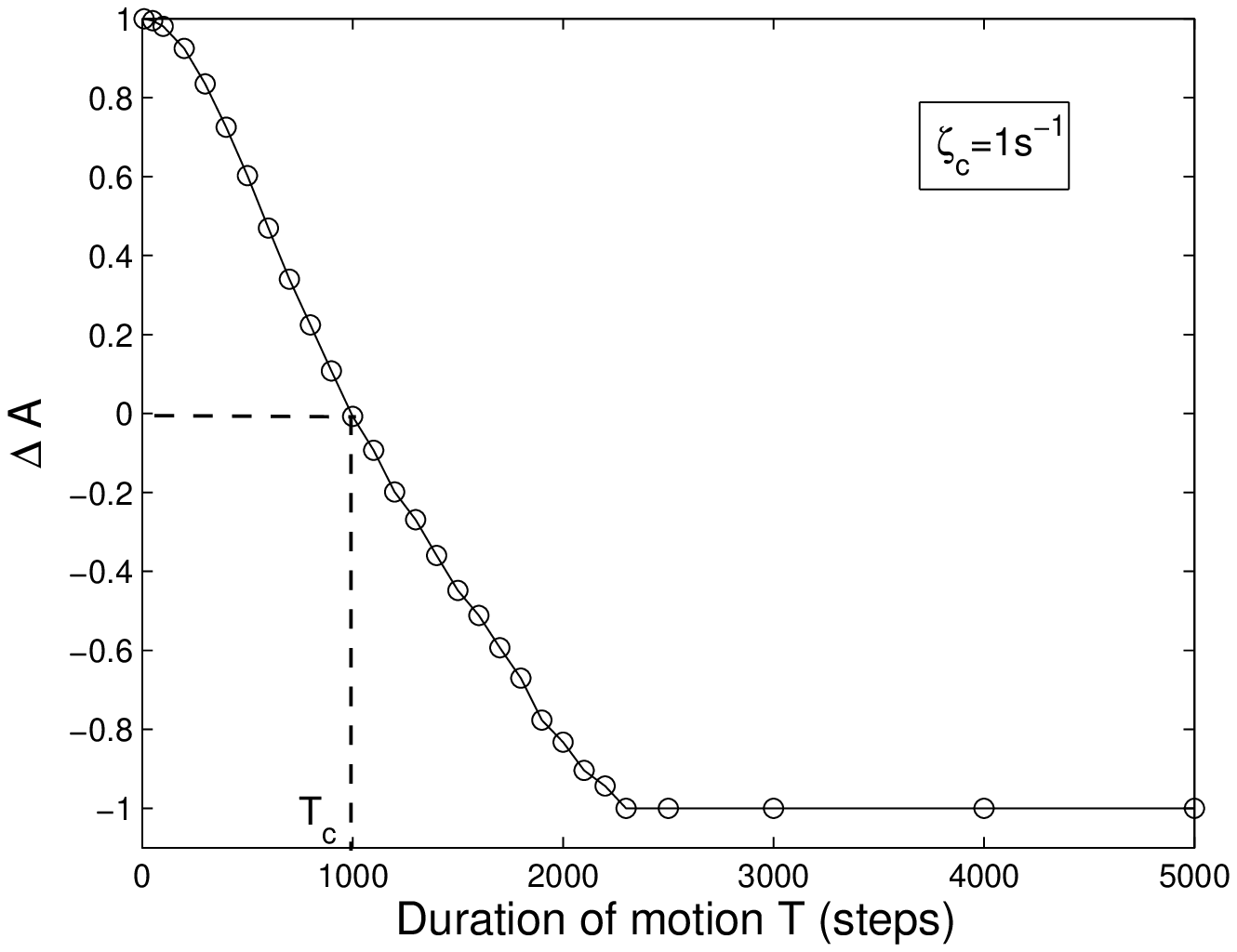}
  \caption{$T$ dependence of the quantity $\Delta A$ for $\zeta=1$ $s^{-1}$ and $\delta t =10^{-3}$ $s$. The characteristic point $T_c$ of the evolution can be approximated by $T_c=1/\zeta$.}
  \label{Fig7}
\end{figure}

For given $\zeta$, the evolution of extrema $\delta A=0$ is a function of the duration of motion $T$. The characteristic point $T_c$ for $\Delta A=0$ can be approximately determined with $\zeta T_c=1$, as shown in Fig.\ref{Fig7} which reveals that the three regimes of the evolution of extrema can be characterized by $\zeta T <<1$ (minimum regime), $\zeta T >>1$ (maximum regime), and  $\zeta T \approx 1$ (saddle point regime).

Further study with different $T$, $\zeta$ and $\sigma$ revealed that this evolution of extrema begins by the lose of the least action whenever $\zeta$ is different from zero. For arbitrarily small $\zeta$, we could always find a $\sigma$ sufficiently small to create paths having smaller actions than $A_{op}$ of the optimal path. For example, Fig.\ref{Fig5} (a) was created with $\zeta=0.1$ $s^{-1}$ and $\sigma=0.1$ $mm$. If we use $\sigma=1$ $nm$, other circles below the dot will appear. In other words, the least action $\delta A_0=0$ for non dissipative systems is definitely lost whenever $A_d$ and $\delta A_d$ are nonzero. However, very small $\sigma$ produces so small perturbations of the optimal path and all the perturbed paths are so close to each other that they can be considered as in the set of the optimal paths. Therefore, from practical point of view, for very small $\zeta$, $\delta A\approx\delta A_0=0$ is a minimum and determines the set of optimal paths which have the smallest actions among all the possible paths with arbitrary perturbations.

Similar discussion can be made for the maximum regime illustrated in Fig.\ref{Fig5} (c). For arbitrarily large $\zeta$ ($10^{10}$ $s^{-1}$ for instance), we could always find sufficiently small $\sigma$ ($10^{-10}$ $m$ for instance) to create paths having larger action than $A_{op}$ (circles above the dot). But these paths are so close to the optimal one that they can be considered as the set (bundle) of paths having the largest action. In this sense, it is sure that $\delta A=0$ is a maximum for large $\zeta$ or overdamped motion. 

\section{Other forces}

From the above results, it is clear that the transition of extrema of action from minimum to maximum is caused by the increasing dissipative energy $E_d$ or its time integral $A_d$. In principle, whenever $A_d$ is no more negligible with respect to $A_0$, the minimum action is lost, and when $A_d$ approaches $A_0$, the maximum action occurs as can be seen from Fig.\ref{Fig2}, \ref{Fig3} and \ref{Fig5}.

\begin{figure}[ht]
\centering
\begin{minipage}[t]{0.49\linewidth}
  \includegraphics[width=0.9\linewidth]{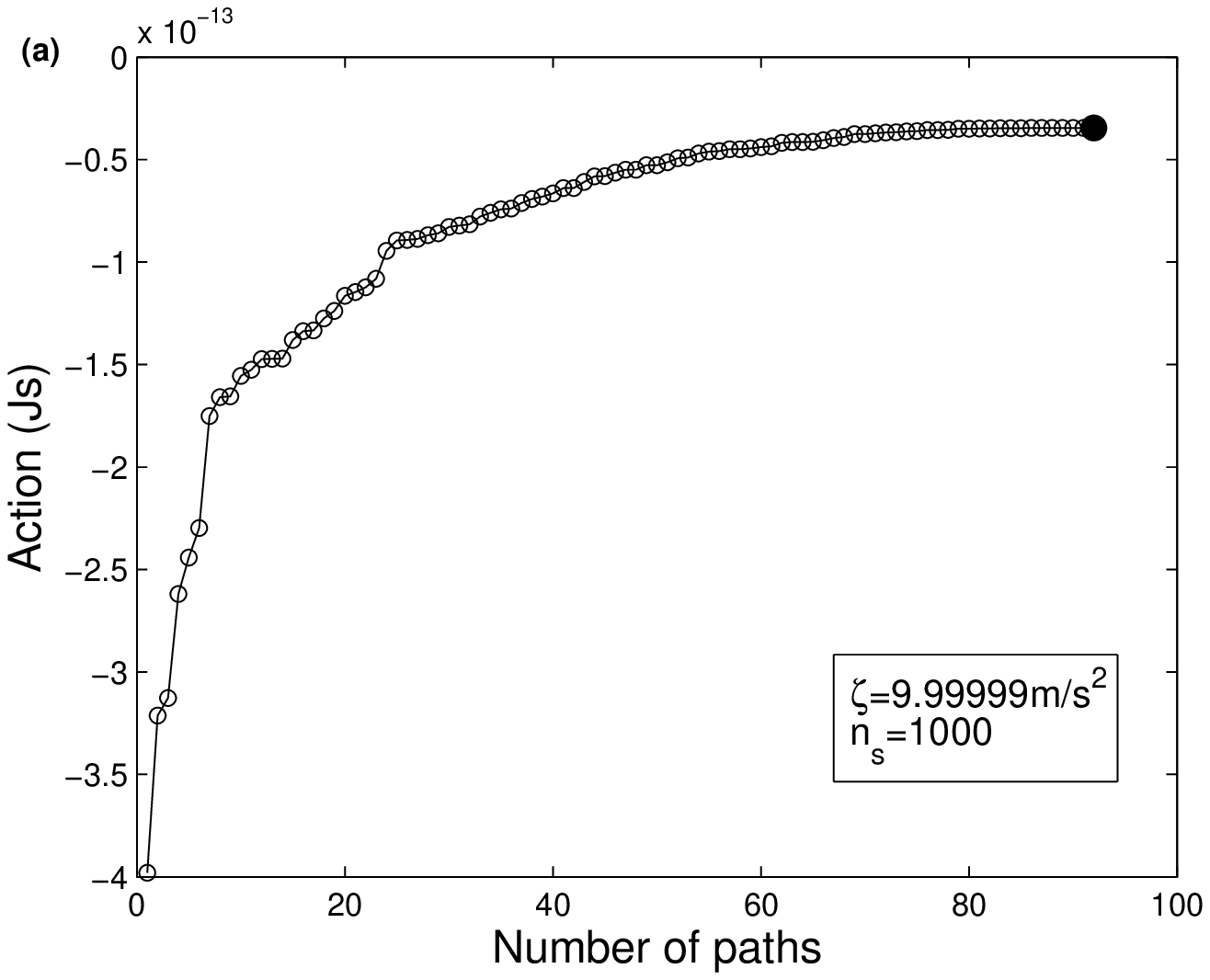}
  \centering
\end{minipage}
\begin{minipage}[t]{0.49\linewidth}
  \includegraphics[width=0.9\linewidth]{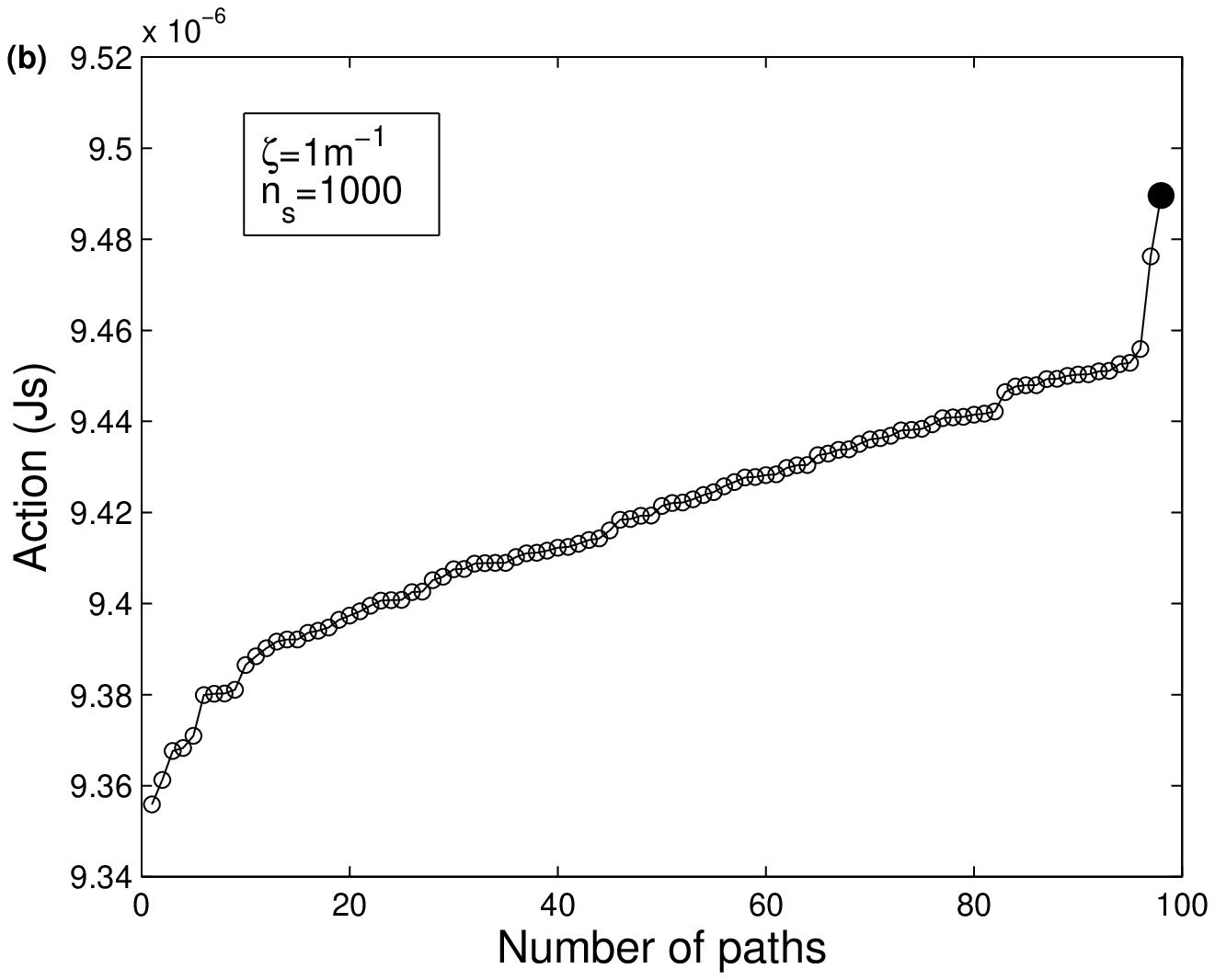}
  \centering
\end{minipage}
\begin{minipage}[t]{0.49\linewidth}
  \includegraphics[width=0.9\linewidth]{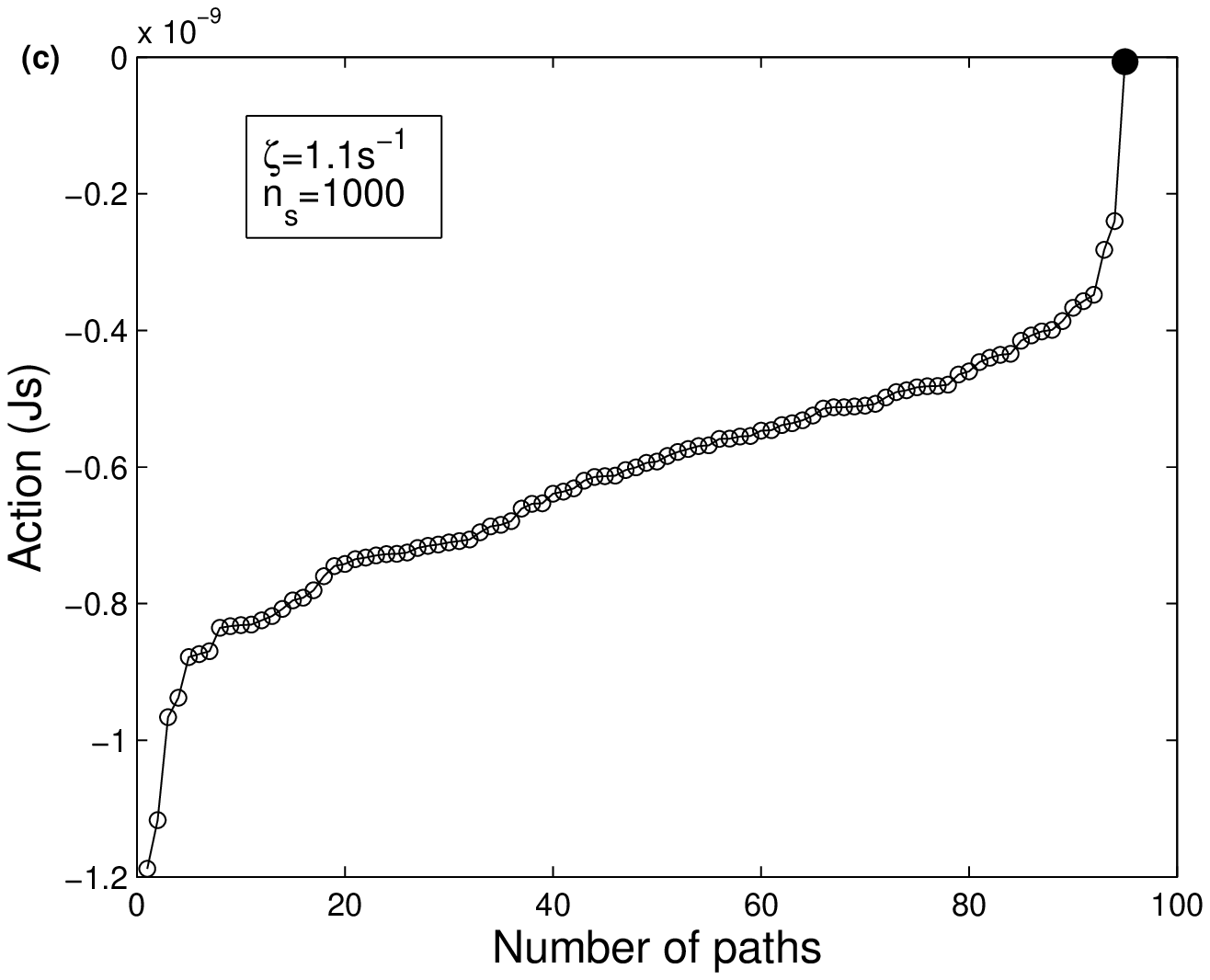}
  \centering
\end{minipage}  
\caption{Comparison of the action of the optimal path (dot) with the actions of the perturbed paths (circles) for 3 overdamped motions : (a) with constant conservative force damped by constant friction $f_d=m\zeta$, where $\zeta=9.99$ $ms^{-2}$ and $\sigma=0.1$ $nm$; (b) with constant conservative force damped by the quadratic drag $f_d=m\zeta \dot{x}^2$, where $\zeta=1$ $m^{-1}$ and $\sigma=0.1$ $mm$;  and (c) with harmonic oscillator damped by Stokes' drag, where $\zeta=1.1$ $s^{-1}$ and $\sigma=0.1$ $mm$. The number of steps is $n_s=1000$ with $\delta t=10^{-3}$ $s$ each step ($T=1$ $s$). In all three cases, the action of the Newtonian path (dot) is a maximum while it was a minimum for small $\zeta$ (not shown here). The $\zeta$-dependent transition of extrema is similar to Fig.5.}
  \label{Fig8}
\end{figure}

From this point of view, similar transition of extrema of action is expected for other friction and conservative forces. We have made simulations for the above system (subject to constant conservative force) damped by constant friction and the quadratic drag $f_d=m\zeta \dot{x}^2$, as well as for harmonic oscillator damped by Stokes' drag. All these cases have similar evolution of extrema from minimum to maximum as shown in Fig.\ref{Fig5} (a), (b) and (c), i.e., the optimal action $A_{op}$ undergoes a transition from minimum to maximum as the motion goes from underdamped state to overdamped state. The maxima of the optimal action in the overdamped regime are shown in Fig.\ref{Fig8}.

\section{Concluding remarks}

The question we want to answer in this work is whether or not an action, defined with the Lagrangian $L=K-V-E_d$ for dissipative systems where $E_d$ is the energy dissipated by friction, is stationary for the path of Newtonian equation of motion. The context is the absence of definitive answer from the calculus of variation, a consequence of the non locality of $E_d$ depending on the history of the motion. The main part of this work is a study of the stationarity of the action by numerical simulation of the damped motion along the optimal path (solution of Newtonian equation) and along many perturbed paths around the optimal one. These perturbed paths can be considered as the paths with arbitrary variations. In other words, this is a numerical simulation of variational calculus to show the possible minima, maxima and inflection points of that action. 

The simulation is made for a particle subject to constant conservative force and harmonic force, combined with three friction forces: constant friction, Stokes' drag and quadratic drag. The comparison of the action values of different paths reveals that the minimum action $A_{op}$ of the optimal path persists in the case of weak dissipation, and is replaced by maximum action in the case of strong dissipation. Hence the extrema $\delta A_{op}=0$ of the optimal path in the underdamped and overdamped cases are confirmed by the numerical simulation. More precisely, when the dissipative energy is negligible (underdamping), the family of the Newtonian optimal paths (a sufficiently thin tube containing the solution of the Newtonian equation) has, as expected, the smallest action. When the dissipative energy is large enough (overdamping), this family of paths has the largest action. 

However, there is no clear evidence for the stationarity $\delta A_{op}=0$ in the intermediate case, the simulation result showing only that $A_{op}$ is neither a minimum nor a maximum. Its rank among all the calculated actions shifts from minimum to maximum with increasing dissipated energy. It is expected that, for $A_{op}$ to be a inflection point, the slope of the curve in the vicinity of the dot (optimal path) in Fig.\ref{Fig5} (b) is vanishing with very small $\sigma$, i.e., the action variation is vanishing when the perturbation of the optimal path is vanishingly small. We have checked this slope with very small variation such as, for instance, $\sigma=10^{-10}$ m for a path of several meters in length. But the slope is not vanishing. However, the inflection point of stationary action in the case of intermediate friction is supported by the following reasoning on the basis of the action extrema in the underdamped and overdamped cases. These extrema implies that, to derive the Newtonian equation from LAP $\delta A_{op}=0$, it is necessary not to consider the variation $\delta x(\tau)$ in the variational calculus (see appendix). In this case, the optimal path with intermediate friction, as a solution of the Newtonian equation, necessarily has $\delta A_{op}=0$ corresponding to an inflection point of action. In any case, the verification of this action inflection by numerical simulation is one of the matters of investigation in the future. 

This work is only carried out with two conservative forces and three frictions. Hence the conclusion of this work is restrictive. To confirm completely the least action principle with the new action for dissipative systems, it will be necessary to check it with variational calculus and to confirm it with as many conservative and frictional forces as possible. Concerning the calculus of variation presented in the appendix, we think that some constraints of variation may be necessary for it to be useful.

Finally, it is worth noticing that the transition of the action stationarity from minimum to maximum with increasing dissipation is a specific behavior of the motions considered in the present work. The reason is that the optimal paths of these motions all have minimum action when there is no dissipation ($\zeta=0$). It will be interesting to see the evolution of the action stationarity when the action of the optimal path with zero dissipation is a maximum or an inflection, cases frequently observed in many mechanical systems\cite{Gray3}.

\section*{Acknowledgements}
This work was supported by the Region des Pays de la Loire in France under the grants No. 2007-6088 and No. 2010-11967.

\section*{Appendix}

The dissipative action being defined by $A=\int_0^T (K-V-E_d)dt$ with $E_d=\int_0^{x(t)} f_d(\tau)dx(\tau)$ or $E_d=\int_0^t f(\tau)d\tau$ and $f=f_d(\tau)\dot{x}(\tau)$. The ``global'' variational calculus, illustrated in Fig.\ref{Fig1}, which consists in considering both the variation $\delta x(t)$ and the antecedent $\delta x(\tau)$, is given by
$$\delta A=\int_0^{T}\delta(K-V-E_d)dt,$$
where the first part will be denoted by $\delta A_0=\int_0^{T}\delta(K-V)dt$ (without dissipative energy) which is

$$\delta A_0=\int_0^{T}\left[\frac{\partial (K-V)}{\partial  x(t)}\delta x(t)+\frac{\partial (K-V)}{\partial \dot{x}(t)}\delta\dot{x}(t) \right]dt$$
in which the variations $\delta x(t)$ and $\delta\dot{x}(t)$ only take place at time $t$. Integrating the second term by parts and using the boundary conditions $\delta x(0)=\delta x(T)=0$, we get

$$\delta A_0=\frac{\partial (K-V)}{\partial \dot{x}(t)}\delta x(t)\bigg |^{T}_{0}+\int_0^{T} \left[\frac{\partial (K-V)}{\partial x(t)}-\frac{d}{dt}\left(\frac{\partial (K-V)}{\partial \dot{x}(t)}\right)\right]\delta x(t)dt$$
$$=\int_0^{T} \left[\frac{\partial (K-V)}{\partial x(t)}-\frac{d}{dt}\left(\frac{\partial (K-V)}{\partial \dot{x}(t)}\right)\right]\delta x(t)dt.$$
The second part of the total variation $\delta A$ will be denoted by $\delta A_d=\int_0^{T}\delta E_d dt$ which is given by
$$\delta A_d=\int_0^{T}\int_0^{t} \left[\frac{\partial f}{\partial  x(\tau)}\delta x(\tau)+\frac{\partial f}{\partial \dot{x}(\tau)}\delta\dot{x}(\tau)\right]d\tau dt$$
where the variations $\delta x(\tau)$ and $\delta\dot{x}(\tau)$ take place at time $\tau$. 

Now let us make an integration by parts with respect to $\tau$ of the term containing $\delta\dot{x}(\tau)$. The result is
$$\delta A_d=\int_0^{T}\left[\int_0^{t} \frac{\partial f}{\partial  x(\tau)}\delta x(\tau)d\tau+\frac{\partial f}{\partial \dot{x}(\tau)}\delta x(\tau)\bigg |^{t}_{0}-\int_0^{t}\frac{d}{d\tau}\left(\frac{\partial f}{\partial \dot{x}(\tau)}\right)\delta x(\tau)d\tau \right]dt$$
$$=\int_0^{T} \frac{\partial f}{\partial \dot{x}(\tau)}\delta x(\tau)\bigg |^{t}_{0}dt+\int_0^{T}\int_0^{t} \left[\frac{\partial f}{\partial x(\tau)}-\frac{d}{d\tau}\left(\frac{\partial f}{\partial \dot{x}(\tau)}\right)\right]\delta x(\tau) d\tau dt.$$
Due to the boundary condition $\delta x(0)=0$, the first term is equal to $\int_0^{T} \frac{\partial f}{\partial \dot{x}(t)}\delta x(t)dt$. Making an integration by parts of $\int_0^{t} \left[\frac{\partial f}{\partial x(\tau)}-\frac{d}{d\tau}\left(\frac{\partial f}{\partial \dot{x}(\tau)}\right)\right]\delta x(\tau)d\tau$ with respect to $t$, $\delta A_d^{(2)}$ turns out to be

$$\delta A_d =\int_0^{T} \frac{\partial f}{\partial \dot{x}(t)}\delta x(t)dt+\left\{t\int_0^{t} \left[\frac{\partial f}{\partial x(\tau)}-\frac{d}{d\tau}\left(\frac{\partial f}{\partial \dot{x}(\tau)}\right)\right]\delta x(\tau)d\tau\right\}\bigg |^{T}_{0}$$
$$-\int_0^{T} t \frac{d}{dt} \left\{\int_0^{t} \left[\frac{\partial f}{\partial x(\tau)}-\frac{d}{d \tau}\left(\frac{\partial f}{\partial \dot{x}(\tau)}\right)\right]\delta x(\tau)d\tau \right\}dt$$
$$=\int_0^{T} \frac{\partial f}{\partial \dot{x}(t)}\delta x(t)dt+T \int_0^{T} \left[\frac{\partial f}{\partial x(\tau)}-\frac{d}{d \tau}\left(\frac{\partial f}{\partial \dot{x}(\tau)}\right)\right]\delta x(\tau)d\tau$$
$$-\int_0^{T} t\left[\frac{\partial f}{\partial x(t)}-\frac{d}{dt}\left(\frac{\partial f}{\partial \dot{x}(t)}\right)\right]\delta x(t)dt.$$
Replacing $\tau$ in the second integral by $t$, we get:
$$\delta A_d=\int_0^{T} \left\{\frac{\partial f}{\partial \dot{x}(t)}+(T-t)\left[\frac{\partial f}{\partial x(t)}-\frac{d}{dt}\left(\frac{\partial f}{\partial \dot{x}(t)}\right)\right]\right\}\delta x(t)dt. $$
Finally, $\delta A=\delta A_0-\delta A_d$ is given by
$$\delta A=\int_0^{T} \left\{\frac{\partial (K-V)}{\partial x(t)}-\frac{d}{dt}\left(\frac{\partial (K-V)}{\partial \dot{x}(t)}\right)\right\}\delta x(t)dt$$
$$-\int_0^{T} \left\{\frac{\partial f}{\partial \dot{x}(t)}+(T-t)\left[\frac{\partial f}{\partial x(t)}-\frac{d}{dt}\left(\frac{\partial f}{\partial \dot{x}(t)}\right)\right] \right\}\delta x(t)dt.$$
The least action principle $\delta A=0$ implies :
$$\frac{\partial (K-V)}{\partial x}-\frac{d}{dt}\left(\frac{\partial (K-V)}{\partial \dot{x}}\right)-\frac{\partial f}{\partial \dot{x}}-(T-t)\left(\frac{\partial f}{\partial x}-\frac{d}{dt}(\frac{\partial f}{\partial\dot{x}})\right)=0.$$
This equation is not a correct Newtonian equation of motion. For example, with Stokes' drag $f_d=m\zeta \dot{x}$, the above equation becomes
$$m\ddot{x}=-\frac{\partial V}{\partial x}-2m\zeta\dot{x}+2(T-t)m\zeta\ddot{x}.$$
This is not the expected equation of motion Eq.(\ref{e5}). The extra terms are
$$-m\zeta\dot{x}+2(T-t)m\zeta\ddot{x}$$
which are not vanishing, in general, for arbitrary dissipation $\zeta$ and duration of motion $T$. 

It should be noticed that the Newtonian equation is simply
$\frac{\partial (K-V)}{\partial x}-\frac{d}{dt}\left(\frac{\partial (K-V)}{\partial \dot{x}}\right)-f_d=0$, and that, from the above calculus, all the extra terms
$$f_d-\frac{\partial f}{\partial \dot{x}}-(T-t)\left(\frac{\partial f}{\partial x}-\frac{d}{dt}(\frac{\partial f}{\partial\dot{x}})\right)$$
come from the consideration of the variations $\delta x(\tau)$ and $\delta\dot{x}(\tau)$ at time $\tau$ prior to the current moment $t$ of the motion. If only $\delta x(t)$ and $\delta\dot{x}(t)$ are considered, the Newtonian equation will be a consequence of $\delta A=0$.


\begin{thebibliography}{99}

\bibitem {Maupertuis}
P.L.M. de Maupertuis, Essai de cosmologie (Amsterdam, 1750) ; Accord de diff\'erentes lois de la nature qui avaient jusqu'ici paru incompatibles. (1744), M\'em. As. Sc. Paris p. 417; Les lois de mouvement et du repos, d\'eduites d'un principe de m\'etaphysique. (1746) M\'em. Ac. Berlin, p. 267

\bibitem {Arnold}
V.I. Arnold, Mathematical methods of classical mechanics, second edition, Springer-Verlag, New York, 1989

\bibitem {Lanczos}
C. Lanczos, The variational principles of mechanics, Dover Publication, New York (1986)

\bibitem {Goldstein}
H. Goldstein, Classical Mechanics, 2nd ed. Reading, Mass.: Addison-Wesley (1981)

\bibitem {Sieniutycz}
S. Sieniutycz and H. Farkas, Variational and extremum principles in macroscopic systems, Elsevier, 2005

\bibitem {Vujanovic}
B.D. Vujanovic and S.E. Jones, Variational methods in nonconservative phenomena, Academic Press Inc., New York, 1989

\bibitem {Goldstine}
H.H. Goldstine, A history of the calculus of variations from the 17th through the 19th century, Springer-Verlag, New York, 1980

\bibitem {Gray}
C.G. Gray, Principle of least action, Scolarpedia, {\bf 4}(2009)8291

\bibitem {Gray2}
C.G. Gray, G. Karl and V.A.Novikov, Progress in Classical and Quantum Variational
Principles, Reports on Progress in Physics, {\bf 67}(2004)159 

\bibitem {Herrera}
L. Herrera, L. Nunez, A. Patino, H. Rago, A veriational principle and the classical and quantum mechanics of the damped harmonic oscillator, Am. J. Phys., {\bf 54}(1986)273

\bibitem {Bateman}
H. Bateman, On dissipative systems and related variational principles, Physical Review {\bf 38}(1931)815

\bibitem {Sanjuan}
M A F Sanjuan, Comments on the Hamiltonian formulation for linear and nonlinear oscillators including dissipation, Journal of Sound and Vibration {\bf 185}(1995)734

\bibitem {Riewe}
F. Riewe, Mechanics with fractional derivatives, Physical Review E {\bf 55}(1997)3581.

\bibitem {Duffin}
R. J. Duffin, Arch. Rat. Mech. Anal. {\bf 9} (1962)309

\bibitem {Schuch}
D. Schuch, A New Lagrange-Hamilton Formalism for Dissipative Systems, International Journal of Quantum Chemistry: Quantum Chemistry Symposium {\bf 24}, 767-780 (1990) 

\bibitem {Wang}
Q.A. Wang, R. Wang, Is it possible to formulate least action principle for dissipative systems? submitted, arXiv:1201.6309


\bibitem {Onsager}
L. Onsager, Reciprocal relations in irreversible processes, Phys. Rev., {\bf 37}(1931)405

\bibitem {Liu}
Y. Hyon, D. Kwak, C. Liu, Energetic variational approach in complex fluids: Maximum dissipation principle, Discrete And Continuous Dynamical Systems, {\bf 26}(2009)1291-1304 

\bibitem {Maupertuiscalculus}

The usual variational calculus with Maupertuis action is as follows
$$
\delta A_M=\delta\int_a^b pdx=\int_a^b(\delta pdx+p\delta dx).
$$
Substitute $dx=\dot{x}dt$ for $dx$, the first term becomes $\delta(\frac{p^2}{2m})dt$ and the second term becomes $-m\ddot{x}\delta xdt$ through the time integration by part of $\delta\dot{x}$ under the condition $\delta x(a)=\delta x(b)=0$. The total energy conservation at the moment $t$ of the variation means $\delta H=\delta(\frac{p^2}{2m})+\frac{\partial V_1}{\partial x}\delta x+\frac{\partial E_d}{\partial x}\delta x=0$ or $\delta(\frac{p^2}{2m})=-\frac{\partial V_1}{\partial x}\delta x-\frac{\partial E_d}{\partial x}\delta x$. Finally, we have
$$
\delta A_M=\int_0^T\left[-\frac{\partial V_1}{\partial x}-\frac{\partial E_d}{\partial x}-m\ddot{x}\right]\delta xdt
$$
which implies that the Maupertuis principle $\delta A_M=0$ necessarily leads to the Newtonian equation of damped motion:
$$
m\ddot{x}=-\frac{\partial (V_1+E_d)}{\partial x}=-\frac{\partial V_1}{\partial x}-f_d
$$
where we have used $$f_d=\frac{\partial E_d}{\partial x}=\frac{\partial}{\partial x}\int_{x_a}^{x}f_d ds$$ according to the second fundamental theorem of calculus (http://mathworld.wolfram.com/SecondFundamentalTheoremofCalculus.html).

\bibitem {Gray3}
C. G. Gray and E. F. Taylor, When Action is Not Least, Am. J. Phys. {\bf 75}(2007)434-458


\end{thebibliography}
\end{document}